\documentclass[prb,preprint,superscriptaddress,showpacs,amsmath,amssymb]{revtex4}

\usepackage{epsfig}

\newcommand{\pdiff}[2]{\frac{\partial#1}{\partial#2}}

\begin{document}

\title{Effective interactions and large-scale diagonalization for quantum dots}
\date{\today}

\author{Simen Kvaal}
\email{simen.kvaal@cma.uio.no}
\affiliation{Centre of Mathematics for Applications, University of
  Oslo, N-0316 Oslo, Norway}
\thanks{This work was supported by the Norwegian Research Council}
\author{Morten \surname{Hjorth-Jensen}}
\affiliation{Department of Physics, University of Oslo, N-0316 Oslo, Norway}
\affiliation{Centre of Mathematics for Applications, University of Oslo, N-0316 Oslo, Norway}
\author{Halvor \surname{M\o ll Nilsen}}
\affiliation{Centre of Mathematics for Applications, University of Oslo, N-0316 Oslo, Norway}


\begin{abstract}
The widely used large-scale diagonalization method using harmonic
oscillator basis functions (an instance of the Rayleigh-Ritz method,
also called a spectral method, configuration-interaction method, or
``exact diagonalization'' method) is systematically analyzed using
results for the convergence of Hermite function series. We apply this
theory to a Hamiltonian for a one-dimensional model of a quantum
dot. The method is shown to converge slowly, and the non-smooth
character of the interaction potential is identified as the main
problem with the chosen basis, while on the other hand its important
advantages are pointed out. An effective interaction obtained by a
similarity transformation is proposed for improving the convergence of the
diagonalization scheme, and numerical experiments are performed to
demonstrate the improvement. Generalizations to more particles and
dimensions are discussed.
\end{abstract}

\maketitle

\section{Introduction}

Large-scale diagonalization is widely used in many areas of physics,
from quantum chemistry\cite{helgaker} to nuclear
physics.\cite{caurier2005} It is also routinely used to obtain spectra
of model quantum dots, see for example
Refs.~\onlinecite{Ezaki1997,Maksym1998,Bruce2000,Creffield1999,Hausler1993,Reimann2000,Rontani2006,Ciftja2006,Jauregui1993,Imamura1999,Tavernier2003,Wensauer2004,Helle2005,Xie2006,Tavernier2006,Gylfadottir2006}. The
method is based on a projection of the model Hamiltonian onto a
finite-dimensional subspace of the many-body Hilbert space in
question, hence the method is an instance of the Rayleigh-Ritz
method.\cite{GouldBook1995} Usually, one takes the stance that the
many-body Hamiltonian is composed of two parts $\hat{H}_0$ and
$\hat{H}_1$, treating the latter as a perturbation of the former,
whose eigenfunctions are assumed to be a basis for the Hilbert
space. This leads to a matrix diagonalization problem, hence the name
of the method. As $\hat{H}_1$ often contains the interaction terms of
the model, ``perturbing'' the electronic configuration states of
$\hat{H}_0$, the method is also called the configuration-interaction
method.  In the limit of an infinite basis, the method is \emph{in
principle} exact, and for this reason it is also called ``exact
diagonalization''.  Usually, however, this method is far from exact,
as $\hat{H}_1$ is rarely a small perturbation (in a sense to be
specified in Sec.~\ref{sec:perturbation-def}) while limited computing
resources yield a tight bound on the number of degrees of freedom
available per particle.

In this work we provide mathematical convergence criteria for
configuration-interaction calculations. More specifically, we address
this problem in the case where $\hat{H}_0$ is a harmonic oscillator
(or h.o.~for short), concentrating on a simple one-dimensional
problem. A common model for a quantum dot is indeed a perturbed
harmonic oscillator, and using h.o.~basis functions is also a common
approach in other fields of many-body physics and partial differential
equations settings in general, as it is also known as the Hermite
spectral method.\cite{Tang1993} When we in the following refer to the
configuration-interaction method, or CI for short, it is assumed that
a h.o.~basis is used.

Studying a one-dimensional problem may seem unduly restrictive, but
will in fact enable us to treat realistic multidimensional problems as
well due to the symmetries of the harmonic oscillator. Moreover, we
choose a worst-case scenario, in which the interaction potential
decays very slowly. We argue that the nature of the perturbation
$\hat{H}_1$, i.e., the non-smooth character of the Coulomb potential
or the trap potential, hampers the convergence properties of the
method.  To circumvent this problem and improve the convergence rate,
we construct an effective two-body interaction via a similarity
transformation. This approach, also using a h.o.~basis, is routinely
used in nuclear
physics,\cite{navratil1998,bruce2,navratil2000a}
where the interactions are of a completely different nature. 

The
effective interaction is defined for a smaller space than the original
Hilbert space, but it reproduces exactly the lowest-lying eigenvalues
of the full Hamiltonian. This can be accomplished by a technique
introduced by Suzuki, Okamoto and
collaborators.\cite{Suzuki1982,suzuki2,Suzuki1995,Suzuki1994}
Approaches based on this philosophy for deriving effective
interactions have been used with great success in the nuclear
many-body problem.\cite{navratil1998,bruce2,navratil2000a} For light
nuclei it provides benchmark calculations of the same quality as
Green's function Monte Carlo methods or other {\em ab initio} methods.
See for example Ref.~\onlinecite{kamada2001} for an extensive
comparison of different methods for computing properties of the
nucleus $^4$He. It was also used in a limited comparative
study of large-scale diagonalization techniques and stochastic
variational methods applied to quantum dots.\cite{kalman1998}

We demonstrate that this approach to the CI method for quantum dots
yields a considerable improvement to the convergence rate. This has
important consequences for studies of the time-development of quantum
dots with two or more electrons, as reliable calculations of the
eigenstates are crucial ingredients in studies of coherence. This is
of particular importance in connection with the construction of
quantum gates based on quantum dots.\cite{Loss1998} Furthermore, the
introduction of an effective interaction allows for studies of
many-electron quantum dots via other many-body methods like
resummation schemes such as various coupled cluster theories as
well. As the effective interaction is defined only within the model
space, systematic and controlled convergence studies of these methods
in terms of the size of this space is possible.

The article is organized as follows: In Sec.~\ref{sec:model} the model
quantum dot Hamiltonian is discussed. In Sec.~\ref{sec:exact-diag} we
discuss the CI method and its numerical properties. Central to this
section are results concerning the convergence of Hermite function
series.\cite{Boyd1984,Hille1939} We also demonstrate the results with
some numerical experiments.

In Sec.~\ref{sec:effective-interactions} we discuss the similarity
transformation technique of Suzuki and
collaborators\cite{Suzuki1982,suzuki2,Suzuki1995,Suzuki1994} and
replace the Coulomb term in our CI calculations with this effective
interaction. We then perform numerical experiments with the new method
and discuss the results.

We conclude the article with generalizations to more particles in
higher dimensions and possible important applications of the new
method in Sec.~\ref{sec:discussion}.

\section{One-dimensional quantum dots}
\label{sec:model}

A widely used model for a quantum dot containing $N$ charged fermions
is a perturbed harmonic oscillator with Hamiltonian 
\begin{eqnarray}
\hat{H} &=& \sum_{j=1}^N \Big( -\frac{1}{2}\nabla_j^2 + \frac{1}{2}\|\vec{r}_j\|^2
+ v(\vec{r}_j) \Big) \nonumber\\*
& & + \sum_{j=1}^N\sum_{k=j+1}^N U(\|r_j-r_k\|),
\label{eq:big-hamiltonian} \end{eqnarray}
where $\vec{r}_j\in\mathbb{R}^2$,
$j=1,\ldots,N$ are each particle's spatial coordinate, $v(\vec{r})$ is
a small modification of the h.o.~potential $\|\vec{r}\|^2/2$, and
$U(r)$ is the Coulomb interaction, viz, $U(r) = \lambda/r$.  Modelling
the quantum dot geometry by a perturbed harmonic oscillator is
justified by self-consistent
calculations,\cite{Kumar1990,Macucci1997,Maksym1997} and is the stance
taken by many other authors using the large-scale diagonalization
technique as
well.\cite{Ezaki1997,Maksym1998,Imamura1999,Bruce2000,Reimann2000,Tavernier2003,Wensauer2004,Helle2005,Ciftja2006,Rontani2006,Xie2006,Tavernier2006}

Electronic structure calculations amount to finding eigenpairs
$(E,\Psi)$, e.g., the ground state energy and wave function, such that
\[ \hat{H}\Psi = E\Psi, \quad \Psi\in\mathcal{H} \text{ and } E\in\mathbb{R}. \]
Here, even though the Hamiltonian only contains spatial coordinates,
the eigenfunction $\Psi$ is a function of both the spatial coordinates
$\vec{r}_k\in \mathbb{R}^2$ and the spin degrees of freedom
$\sigma_k\in\{-1/2,+1/2\}$, i.e.,
\[ \mathcal{H} = L_2(\mathbb{R}^{2N})\otimes \mathbb{C}^2. \]
The actual Hilbert space is the space of the \emph{antisymmetric}
functions, i.e., functions $\Psi$ for which
\[ \Psi(x_{P(1)}, x_{P(2)}, \ldots, x_{P(N)}) =
\operatorname{sgn}(P)\Psi(x_1,x_2,\ldots,x_N), \] 
for all permutations $P$ of $N$ symbols. Here, $x_k = (\vec{r}_k,\sigma_k)$.

For simplicity, we concentrate on one-dimensional quantum
dots. Even though this is not an accurate model for real quantum dots,
it offers several conceptual and numerical advantages. Firstly, the
symmetries of the harmonic oscillator makes the numerical properties
of the configuration-interaction method of this system very similar to
a two or even three-dimensional model, as the analysis extends almost
directly through tensor products. Secondly, we may investigate
many-body effects for moderate particle numbers $N$ while still
allowing a sufficient number of h.o.~basis functions for unambiguously
addressing accuracy and convergence issues in numerical experiments.

In this article, we further focus on two-particle quantum
dots. Incidentally, for the two-particle case one can show that the
Hilbert space of anti-symmetric functions is spanned by functions on
the form
\[ 
  \Psi(\vec{r}_1,\sigma_1,\vec{r}_2,\sigma_2) =
  \psi(\vec{r}_1,\vec{r}_2)\chi(\sigma_1,\sigma_2),
\]
where the spin wave function $\chi$ can be taken as symmetric or
antisymmetric with respect to particle exchange, leading to an
antisymmetric or symmetric spatial wave function $\psi$,
respectively. Inclusion of a magnetic field $\vec{B}$ poses no
additional complications,\cite{Wensauer2003} but for simplicity we
presently omit it. Thus, it is sufficient to consider the spatial
problem and produce properly symmetrized wavefunctions.

Due to the peculiarities of the bare Coulomb potential in one
dimension\cite{Kurasov1996,Gesztesy1980} we choose a screened
approximation $U(x_1-x_2;\lambda,\delta)$ given by
\[ U(x;\lambda,\delta) = \frac{\lambda}{|x| + \delta}, \]
where $\lambda$ is the strength of the interaction and $\delta>0$ is a
screening parameter which can be interpreted as the width of the wave
function orthogonal to the axis of motion. This choice is made since
it is non-smooth, like the bare Coulomb potential in two and three
dimensions.  The total Hamiltonian then reads
\begin{eqnarray}
  \hat{H} &=&
  -\frac{1}{2}\Big(\pdiff{^2}{x_1^2}+\pdiff{^2}{x_2^2}\Big) +\frac{1}{2}(x_1^2+x_2^2) + \nonumber \\*
  && v(x_1)+v(x_2)+ U(x_1-x_2;\lambda,\delta). \label{eq:ham2} 
\end{eqnarray} 

Observe that for $U=0$, i.e.,
$\lambda=0$, the Hamiltonian is separable. The eigenfunctions of
$\hat{H}$ (disregarding proper symmetrization due to the Pauli
principle) become $\psi_{n_1}(x_1)\psi_{n_2}(x_2)$, where $\psi_n(x)$
are the eigenfunctions of the trap Hamiltonian $\hat{H}_\text{t}$
given by
\begin{equation}\hat{H}_\text{t} = -\frac{1}{2}\pdiff{^2}{x^2} +
\frac{1}{2}x^2 + v(x). \label{eq:trap-ham} \end{equation} Similarly, for a
vanishing trap modification $v(x)=0$ the Hamiltonian is separable in
(normalized) centre-of-mass coordinates given by
\[ X=\frac{x_1+x_2}{\sqrt{2}} \quad\text{and}\quad x=\frac{x_1-x_2}{\sqrt{2}}.\]
Indeed, any orthogonal coordinate change leaves the h.o.~Hamiltonian invariant
(see Sec.~\ref{sec:exact-diag}), and hence
\begin{eqnarray*}
\hat{H} &=&
-\frac{1}{2}\Big(\pdiff{^2}{X^2}+\pdiff{^2}{x^2}\Big) +\frac{1}{2}(X^2+x^2) +\\*
& & v\big((X+x)/\sqrt{2}\big) +
v\big((X-x)/\sqrt{2}\big) + U(\sqrt{2}x;\lambda,\delta).
\end{eqnarray*}
The eigenfunctions become $\phi_n(X)\psi_m(x)$, where $\phi_n(X)$ are the
Hermite functions, i.e., the eigenfunctions of the h.o.~Hamiltonian
(see Sec.~\ref{sec:exact-diag}), and where $\psi_m(x)$ are the
eigenfunctions of the interaction Hamiltonian, viz,
\begin{equation}\hat{H}_\text{i} = -\frac{1}{2}\pdiff{^2}{x^2} + \frac{1}{2}x^2 +
U(\sqrt{2}x;\lambda,\delta). \label{eq:inter-ham} \end{equation}
Odd (even) functions $\psi_m(x)$ yield antisymmetric (symmetric) wave
functions with respect to particle interchange.

\section{Configuration-interaction method}
\label{sec:exact-diag}

\subsection{The harmonic oscillator and model spaces}\label{subsec:modelspaces}

The configuration-interaction method is an
instance of the Rayleigh-Ritz method,\cite{GouldBook1995} employing
eigenfunctions of the unperturbed h.o.~Hamiltonian as basis for a
finite dimensional Hilbert space $\mathcal{P}$, called the model
space, onto which the Hamiltonian \eqref{eq:big-hamiltonian}, or in
our simplified case, the Hamiltonian \eqref{eq:ham2}, is projected and
then diagonalized. As mentioned in the Introduction, this method is
\emph{in principle} exact, if the basis is large enough.

We write the $N$-body Hamiltonian \eqref{eq:big-hamiltonian} as
\[ \hat{H} = \hat{H}_0 + \hat{H}_1, \]
with $\hat{H}_0$ being the h.o.~Hamiltonian, viz,
\[ \hat{H}_0 = -\frac{1}{2}\sum_{j=1}^N \nabla^2_j +
\frac{1}{2}\sum_{j=1}^N \| \vec{r}_j \|^2, \]
and $\hat{H}_1$ being a perturbation of $\hat{H}_0$, viz,
\[ \hat{H}_1 =\sum_{j=1}^N v(\vec{r}_j) + \sum_{j=1}^N\sum_{k=j+1}^N
U(\|r_j-r_k\|). \]
For a simple one-dimensional model of two particles
we obtain
\[ \hat{H}_0 = \hat{h}(x_1) + \hat{h}(x_2), \]
where $\hat{h}(x)$ is the well-known one-dimensional harmonic oscillator
Hamiltonian, viz,
\[ \hat{h}(x) = -\frac{1}{2}\pdiff{^2}{x^2} + \frac{1}{2}x^2. \]
Clearly, $\hat{H}_0$ is just a two-dimensional h.o.~Hamiltonian, if we
disregard symmetrization due to the Pauli principle.
For the perturbation, we have
\[ \hat{H}_1 = v(x_1) + v(x_2) + \frac{\lambda}{|x_1-x_2| +
  \delta}. \]

In order to do a more general treatment, let us recall some basic facts about the harmonic
oscillator.

If we consider a single particle in $D$-dimensional space,
it is clear that the $D$-dimensional harmonic oscillator
Hamiltonian is the sum of one-dimensional h.o.~Hamiltonians for each
Euclidean coordinate, viz,
\begin{equation} 
  \hat{h}^{(D)} = -\frac{1}{2}\nabla^2 +
  \frac{1}{2}\|\vec{x}\|^2 = \sum_{k=1}^D
  \hat{h}(x_k). \label{eq:ho-ham} 
\end{equation}
We indicate the variables on which the operators depend by parenthesis
if there is danger of confusion.  Moreover, the h.o.~Hamiltonian for
$N$ (distinguishable) particles in $d$ dimensions is simply
$\hat{h}^{(Nd)}$. The $D$-dimensional h.o.~Hamiltonian is manifestly
separable, and the eigenfunctions are
\[ \Phi_{\vec{n}}(\vec{x}) = \prod_{k=1}^D \phi_{n_k}(x_k) \]
with energies
\[ \epsilon_{\vec{n}} = \frac{D}{2} + \sum_{k=1}^D n_k,
\]
where $\vec{n}$ denotes the multi-index of quantum numbers
$n_k$. The one-dimensional h.o.~eigenfunctions are given by
\[ \phi_n(x) = \big(2^n n! \pi^{1/2}\big)^{-1/2} H_n(x)e^{-x^2/2}, \]
where $H_n(x)$ are the usual Hermite polynomials. These functions are
the Hermite functions and are treated in further detail in
Sec.~\ref{sec:hermite-series}.

As for the discretization of the Hilbert space, we employ a so-called
\emph{energy-cut model space} $\mathcal{P}$, defined by the
span of all h.o.~eigenfunctions with energies up to a given
$\epsilon = N_\text{max} + D/2$, viz,
\[ \mathcal{P} := \operatorname{sp}\big\{
\Phi_{\vec{n}}(\vec{x}) \; \big| \; 0\leq\sum_k
n_k \leq N_\text{max} \big\}, \]
where we bear in mind that the $D=Nd$ dimensions are distributed among
the $N$ particles.

\begin{figure}[hbpt]
\includegraphics{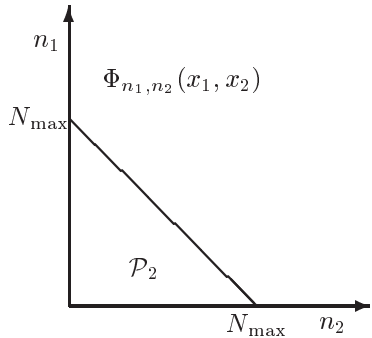}%
 \caption{Two-body model space defined by a cut in energy. The two-body
     state has  quantum numbers $n_1$ and $n_2$, the sum of which does
     not exceed $N_\text{max}$.\label{fig:simtransfP}}
\end{figure}

For the one-dimensional model with only one particle, the model space reduces to 
\begin{equation}\mathcal{P}_1 = \operatorname{sp} \big\{ \phi_n(x) \; \big| \; 0\leq n \leq
N_\text{max}  \big\}. 
\label{eq:one-dim-model-space} \end{equation}
Thus, one particle is associated with one integer quantum number $n$,
denoting the ``shell number where the particle resides'', in typical
terms. For two particles, we get 
\[ \mathcal{P}_2 = \operatorname{sp} \big\{
\phi_{n_1}(x_1)\phi_{n_2}(x_2) \; \big| \; 0\leq n_1 + n_2 \leq N_\text{max}
\big\} . \] 
We illustrate this space in Fig.~\ref{fig:simtransfP}.

Proper symmetrization must also be applied. However, the Hamiltonian
\eqref{eq:big-hamiltonian} commutes with particle permutations,
meaning that the eigenfunctions \emph{will} be symmetric or
antisymmetric, assuming that the eigenvalues are distinct. In the case
of degeneracy, we may simply produce (anti)symmetric eigenfunctions by
taking linear combinations.

We mention that other model spaces can also be used; most common is
perhaps the \emph{direct product model space}, defined by $N$ direct
products of $\mathcal{P}_1$ rather than a cut in energy as above.

\subsection{Coordinate changes and the h.o.}

It is obvious that any orthogonal coordinate change $\vec{y}=S\vec{x}$
where $S^TS=1$ commutes with $\hat{h}^{(D)}$. In particular,
energy is conserved under the coordinate change. Therefore, the
eigenfunctions of the transformed Hamiltonian will be a linear
combination of the original eigenfunctions of the same energy, viz,
\[ \Phi_{\vec{n}}(S\vec{x}) = \sum_{\vec{n}'}
\langle \Phi_{\vec{n}'}, \hat{T}
\Phi_{\vec{n}} \rangle \Phi_{\vec{n}'}(\vec{x}), \] where the sum is
over all $\vec{n}'$ such that
$\epsilon_{\vec{n}'}=\epsilon_{\vec{n}}$. Here, $\hat{T}$ performs the
coordinate change, viz,
\begin{equation}
 \hat{T}\Phi_{\vec{n}}(\vec{x}) = \Phi_{\vec{n}}(S\vec{x}),
\label{eq:trafo}
\end{equation}
where $\hat{T}$ is unitary.  Also note that energy conservation
implies that $\mathcal{P}$ is invariant with respect to the coordinate
change, implying that the CI method is equivalent in the two
coordinate systems.

An important example is the centre-of-mass transformation introduced
in Sec.~\ref{sec:model}. This transformation is essential when we want
to compute the Hamiltonian matrix since the interaction is given in
terms of these coordinates. 

Observe that in the case when the Hamiltonian is in fact separated by
such a coordinate change, the formulation of the \emph{exact} problem using
h.o.~basis is equivalent to two one-particle problems using h.o.~basis
in the new coordinates.

\subsection{Approximation properties of the Hermite functions}
\label{sec:approximation}
\label{sec:hermite-series}

In order to understand the accuracy of the CI method, we need to study
the approximation properties of the Hermite functions.  Note that all
the Hermite functions $\phi_{n}(x)$ spanning $L_2(\mathbb{R})$ are
\emph{smooth}. Indeed, they are holomorphic in the entire complex
plane. Any finite linear combination of these will yield another
holomorphic function, so any non-smooth function will be badly
approximated. This simple fact is sadly neglected in the
configuration-interaction literature, and we choose to stress it here:
Even though the Hermite functions are simple to compute and deal with,
arising in a natural way from the consideration of a perturbation of
the h.o.~and obeying a wealth of beautiful relations, they are not
very well suited for computation of functions whose smoothness is less
than infinitely differentiable, or whose decay behaviour for large
$|x|$ is algebraic, i.e., $f(x) = o(|x|^\beta)$ for some
$\beta<0$. Due to the direct product nature of the $N$-body basis
functions, it is clear that these considerations are general, and not
restricted to the one-dimensional one-particle situation.

Consider an expansion $\psi(x) = \sum_{n=0}^\infty c_n \phi_n(x)$ in
Hermite functions of an arbitrary $\psi\in L_2(\mathbb{R})$. The
coefficients are given by
\[ c_n = \langle \phi_n, \psi \rangle = \int_{-\infty}^\infty
\psi(x)\bar{H}_n(x)e^{-x^2/2} \;\mathrm{d} x. \]
Here, $\bar{H}_n(x)=(2^nn!\sqrt{\pi})^{-1/2}H_n(x)$ are the normalized
Hermite polynomials. If $\psi(x)$ is well approximated by the
basis, the coefficients $c_n$ will decay quickly with increasing
$n$. The \emph{least} rate of convergence is a direct consequence of 
\[ \| \psi \|^2 = \sum_{n=0}^\infty |c_n|^2 < \infty, \]
hence we must have
 $|c_n| = o(n^{-1/2})$. (This is \emph{not} a
sufficient condition, however.) With further restrictions on the
behaviour of $\psi(x)$, the decay will be faster. This
is analogous to the faster decay of Fourier coefficients for smoother
functions,\cite{Tveito2002} although for Hermite functions, smoothness
is not the only parameter as we consider an infinite domain. In this
case, another equally important feature is the decay of $\psi(x)$ as $|x|$
grows, which is intuitively clear given that all the Hermite functions decay as
$\exp(-x^2/2)$.

Let us prove this assertion. We give here a simple argument due to
Boyd (Ref.~\onlinecite{Boyd1984}), but we strengthen the result
somewhat.

To this end, assume that $\psi(x)$ has $k$ square integrable
derivatives (in the weak sense) and that $x^m\psi(x)$ is square integrable for
$m=0,1,\ldots,k$. Note that this is a sufficient condition for
\[ a^\dag \psi(x) = \frac{1}{\sqrt{2}}(x\psi(x) - \psi'(x)), \]
and $(a^\dag)^2\psi(x)$ up to $(a^\dag)^k\psi(x)$ to be square
integrable as well. Here, $a^\dag$ and its Hermitian conjugate $a$ are
the well-known ladder operators for the harmonic
oscillator.\cite{Mota2002}

Using integration by parts, the formula for $c_n$ becomes
\begin{eqnarray*}
 c_n && = \int_{-\infty}^\infty \psi(x)\bar{H}_n(x)e^{-x^2/2} \;\mathrm{d} x \\*
&&= -(n+1)^{-1/2}\int_{-\infty}^\infty [a^\dag\psi(x)]\bar{H}_{n+1}(x)e^{-x^2/2}\;\mathrm{d} x,
\end{eqnarray*}
or
\[ c_n = -(n+1)^{-1/2}d_{n+1}, \]
where $d_n$ are the Hermite expansion coefficients of
$a^\dag\psi(x)\in L_2$. Since $\sum |d_n|^2 < \infty$ by assumption,
we obtain
\[ \sum_{n=0}^\infty n|c_n|^2 < \infty, \]
implying
\[ c_n = o(n^{-1}). \]
Repeating this argument $k$ times, we obtain the estimate
\[ c_n = o(n^{-(k+1)/2}). \]

It is clear that if $\psi(x)$ is infinitely differentiable, and if in
addition $\psi(x)$ decays faster than any power of $x$, such as for
example exponentially decaying functions, or functions behaving like
$\exp(-\alpha x^2)$, $c_n$ will decay faster than \emph{any} power of
$1/n$, so-called ``infinite-order convercence,'' or ``spectral
convergence.'' Indeed, Hille (Ref.~\onlinecite{Hille1939}) gives
results for the decay of the Hermite coefficients for a wide class of
functions. The most important for our application being the following:
If $\psi(x)$ decays as $\exp(-\alpha x^2)$, with $\alpha>0$, and if
$\tau>0$ is the distance from the real axis to the nearest pole of
$\psi(x)$ (when considered as a complex function), then
\begin{equation}|c_n| = O(n^{-1/4}e^{-\tau\sqrt{2n+1}}),
\label{eq:strip-analytic} \end{equation} a very rapid decay for even
moderate $\tau$.

An extremely useful property\cite{Boyd1984} of the Hermite functions
is the fact that they are uniformly bounded, viz,
\[ |\phi_n(x)| \leq 0.816, \quad \forall x, n. \]
As a consequence, the \emph{pointwise} error in a truncated series is
almost everywhere bounded by
\[ | \psi(x) - \sum_{n=0}^{N_\text{max}} c_n \phi_n(x) | \leq 0.816
\sum_{n=N_\text{max}+1}^\infty |c_n|. \]
Thus, estimating the error in the expansion amounts to estimating
the sum of the neglected coefficients. If $|c_n|=o(n^\alpha)$,
\[ | \psi(x) - \sum_{n=0}^{N_\text{max}} c_n \phi_n(x) | = o(N_\text{max}^{\alpha+1}), \quad \text{a.e.} \]
For the error in the mean,
\begin{equation}\| \psi(x) - \sum_{n=0}^N c_n \phi_n(x) \| =
O(N_\text{max}^{\alpha+1/2}), \label{eq:mean-error} \end{equation}
as is seen by approximating $\sum_{n=N_\text{max}+1}^\infty |c_n|^2$ by an
integral.

In the above, ``almost everywhere'', or ``a.e.'' for short, refers to
the fact that we do not distinguish between square integrable
functions that differ on a point set of Lebesgue measure
zero. Moreover, there is a subtle distinction between the notations
$O(g(n))$ and $o(g(n))$. For a given function $f$, $f(n)=o(g(n))$ if
$\lim_{n\rightarrow \infty} |f(n)/g(n)| = 0$, while $f(n)=O(g(n))$ if
we have $\lim_{n\rightarrow \infty} |f(n)/g(n)| < \infty$; a slightly
weaker statement.

\subsection{Application to the interaction potential}
\label{sec:interaction-potential}

Let us apply the above results to the eigenproblem for a perturbed
one-dimensional harmonic oscillator, i.e.,
\begin{equation}\psi''(x) = [x^2 + 2f(x) - 2E]\psi(x), \label{eq:perturbed-osc}
\end{equation}
which is also applicable when the two-particle Hamiltonian
\eqref{eq:ham2} is separable, i.e., when $U=0$ or $v=0$.

It is now clear that under the assumption that $f(x)$ is $k$ times
differentiable (in the weak sense), and that $f(x) = o(|x|^2)$ as
$|x|\rightarrow\infty$, the eigenfunctions will be $k+2$ times
(weakly) differentiable and decay as $\exp(-x^2/2)$ for large
$|x|$. Hence, the Hermite expansion coefficients of $\psi(x)$ will
decay as $o(n^\alpha)$, $\alpha = -(k+3)/2$.

If we further assume that $f(x)$ is analytic in a strip of width
$\tau>0$ around the real axis, the same will be true for $\psi(x)$,
such that we can use Eq.~\eqref{eq:strip-analytic} to estimate the
coefficients. 

A word of caution is however at its place.  Although we have argued
that if a given function can be differentiated $k$ times (in the weak
sense) then the coefficients decay as $o(n^\alpha)$, $\alpha=-(k+1)/2$, it may happen
that this decay ``kicks in'' too late to be observable in practical
circumstances. 

Consider for example the following function:
\[ g(x) = \frac{e^{-x^2/2}}{|x|+\delta}, \]
which has exactly one (almost everywhere continuous) derivative and
decays as $\exp(-x^2/2)$. However, the derivative is seen to have a
jump discontinuity of magnitude $2/\delta^2$ at $x=0$. From the
theory, we expect $o(n^{-1})$ decay of the coefficients, but for small
$\delta$ the first derivative is badly approximated, so we expect
to observe only $o(n^{-1/2})$ decay for moderate $n$, due to the fact
that the rate of decay of the coefficients of $g(x)$ are \emph{explicitely}
given in terms of the coefficients of $a^\dag g(x)$.

\begin{figure*}
\includegraphics[width=0.45\textwidth]{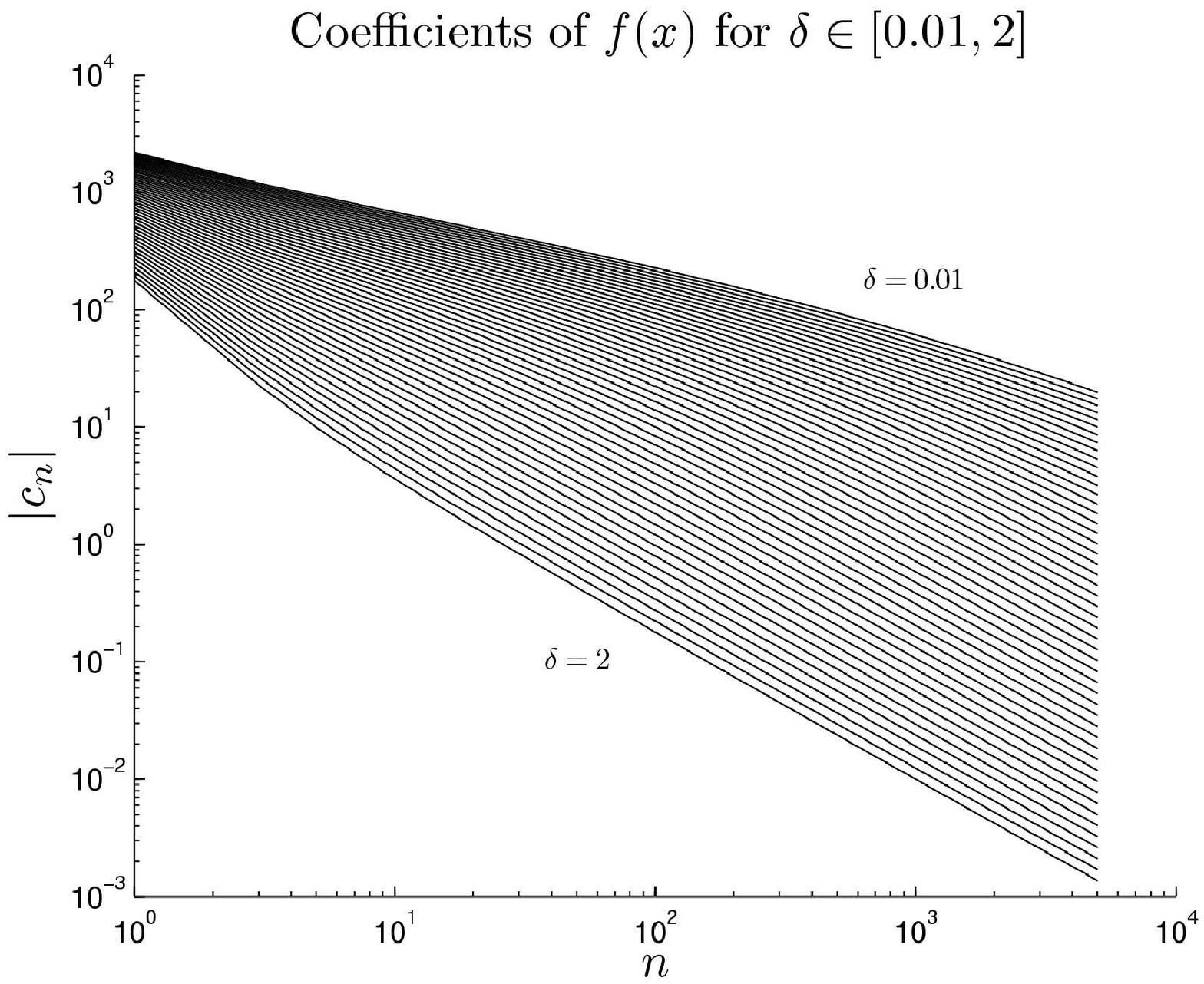}%
\includegraphics[width=0.45\textwidth]{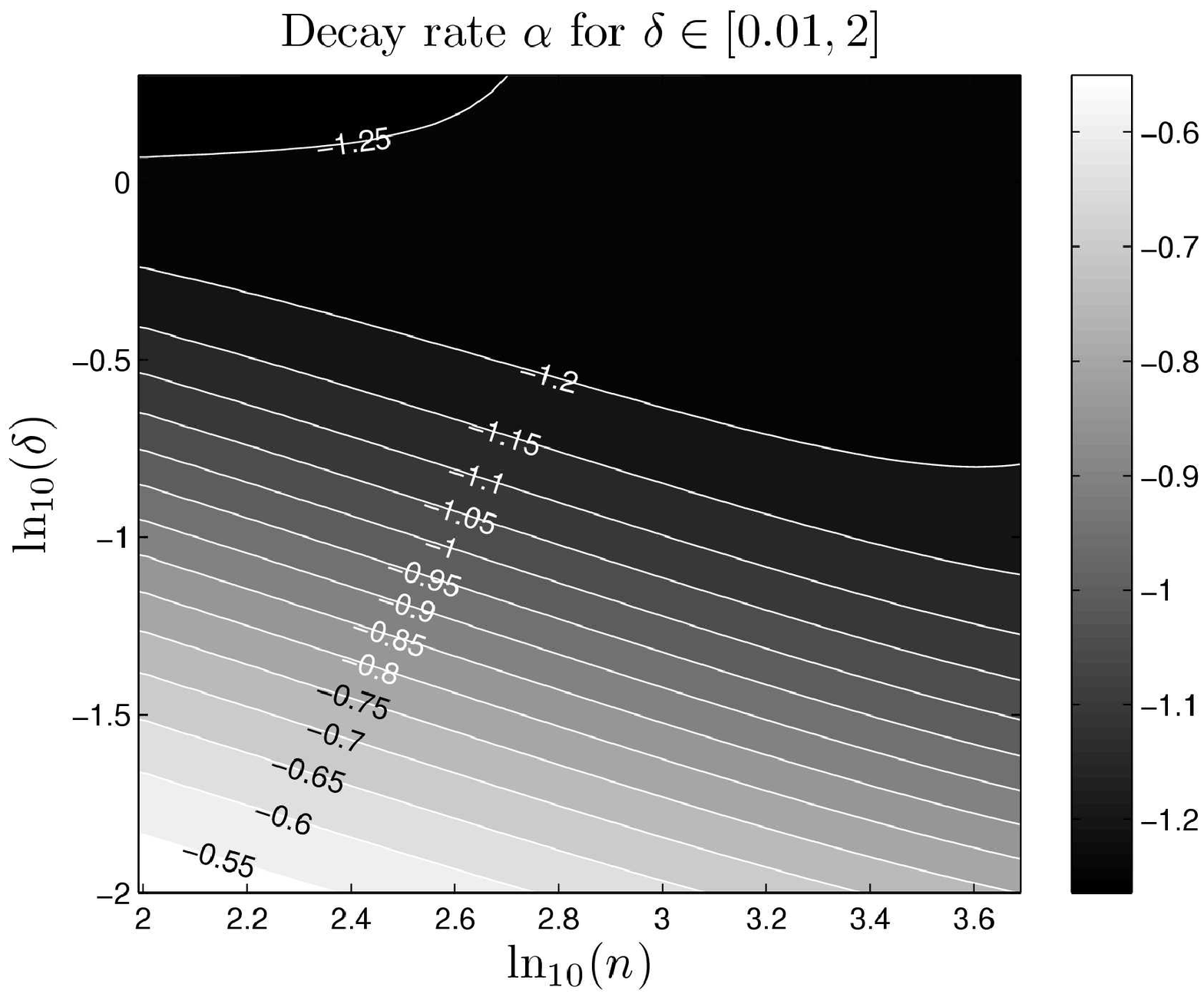}%
\caption{(Left) Coefficients $|c_n|$ of the function
  $\exp(-x^2/2)/(|x|+\delta)$ for $\delta\in[0.01,2]$,
  $n=0,2,\ldots,5000$. (Right) Estimated decay rate $\alpha$ of $|c_n|$, i.e.,
  slope of the graphs in the left panel.\label{fig:bad-func}}
\end{figure*}

In Fig.~\ref{fig:bad-func} the decay rates at different $n$ and for
various $\delta$ are displayed. The decay rate $\alpha$ is computed by
estimating the slope of the graph of $\ln |c_n|$ versus $\ln n$, a
technique used thoughout this article.  Indeed, for small $\delta$ we
observe only $\alpha \approx -1/2$ convergence in practical settings, where
$n$ is moderate, while larger $\delta$ gives $\alpha \approx -1$ even for
small $n$.

The above function was chosen due to its relation to the interaction
Hamiltonian \eqref{eq:inter-ham}. Indeed, its coefficients are given
by
\[ c_n = \langle \phi_n, g \rangle = \langle \phi_n, U(x; 1, \delta)
\phi_0 \rangle, \] i.e., the proportional to the first row of the
interaction matrix. Moreover, due to Eq.~\eqref{eq:perturbed-osc}, the
ground state $\psi$ of the interaction Hamiltonian has a second
derivative with similar behaviour near $x=0$ as $g(x)$. Thus, we
expect to observe $\alpha \approx -3/2$, rather than $\alpha \approx
-2$, for the available range of $n$ in the large-scale diagonalization
experiments.

We remark here, that it is quite common to model quantum dot systems
using non-smooth potentials\cite{Wensauer2000} $v(\vec{r})$, and even
to use the CI method with h.o.~basis functions on these
models.\cite{Harju2002,Helle2005,Forre2006}

\subsection{Numerical experiments}
\label{sec:perturbation-def}

We wish to apply the above analysis by considering the model
Hamiltonian \eqref{eq:ham2}. We first consider the case where $v(x)=0$ or
$U(x)=0$, respectively, which reduces the two-particle problem to
one-dimensional problems through separation of variables, i.e., the
diagonalization of the trap Hamiltonian $\hat{H}_\text{t}$ and the
interaction Hamiltonian $\hat{H}_\text{i}$ in
Eqs.~\eqref{eq:trap-ham} and \eqref{eq:inter-ham}. Then we turn to
the complete non-separable problem.

\begin{figure*}
\includegraphics[width=0.9\textwidth]{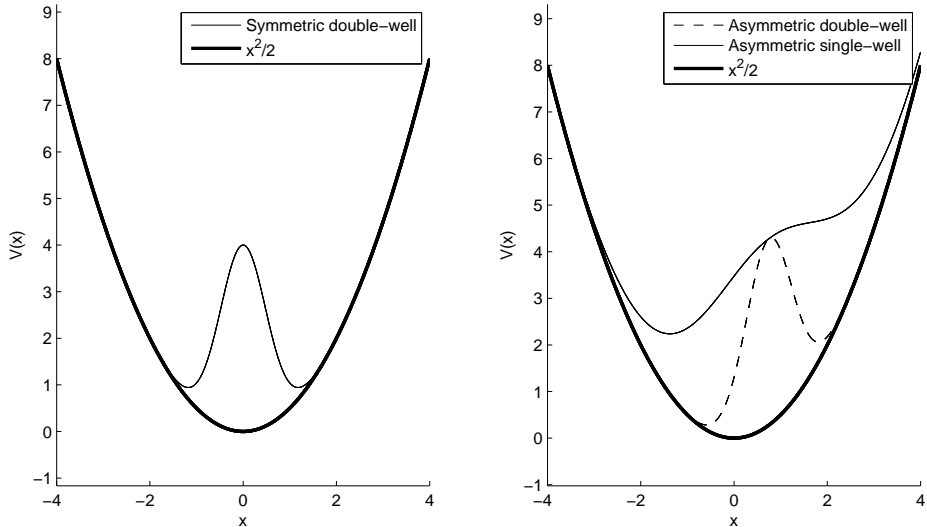}%
\caption{%
Left: Symmetric double-well potential created with the Gaussian
perturbation $A\exp[-C(x-\mu)^2]$ with $A=4$, $\mu=0$ and
$C=2$.
Right: Asymmetric double-well potential created with the Gaussian
perturbation with $A=4$, $\mu=0.75$ and
$C=2$, and single-well potential using $C=0.25$.
\label{fig:potential-discussion}}
\end{figure*}

For simplicity we consider the trap $x^2/2 + v(x)$ with
\[ v(x) = A e^{-C(x-\mu)^2}, \quad A,C>0, \mu\in\mathbb{R},\]
which gives rise to a double-well potential or a single-well
potential, depending on the parameters, as depicted in
Fig.~\ref{fig:potential-discussion}. The perturbation is everywhere
analytic and rapidly decaying. This indicates that
the corresponding configuration-interaction energies and wave functions
also should converge rapidly.  In the below numerical experiments, we
use $A=4$, $C=2$ and $\mu=0.75$, creating the asymmetric double
well in Fig.~\ref{fig:potential-discussion}.

For the interaction Hamiltonian $\hat{H}_\text{i}$ and its potential
$x^2/2 + U(\sqrt{2}x;\lambda,\delta)$ we arbitrarily choose $\lambda =
1$ and $\delta = 0.01$, giving a moderate jump discontinuity in the
derivative.

As these problems are both one-dimensional, the model space reduces to $\mathcal{P}_1$
as given in Eq.~\eqref{eq:one-dim-model-space}. Each problem then amounts
to diagonalizing a matrix $\mathbf{H}$ with elements
\begin{widetext}
\[
\mathbf{H}_{n,m} = \langle \phi_n, \hat{H}_\text{t,i} \phi_m \rangle =
 \big(n+\frac{1}{2}\big)\delta_{n,m} + \int_{-\infty}^\infty
\phi_n(x)f(x)\phi_m(x)\;\mathrm{d} x, \quad 0\leq n,m \leq N_\text{max},
\]
\end{widetext}
with $f(x)=v(x)$ or $f(x)=U(\sqrt{2}x;1,0.01)$. We compute the matrix
to desired precision using Gauss-Legendre quadrature.  In order to
obtain reference eigenfunctions and eigenvalues we use a constant
reference potential method\cite{Ledoux2004} implemented in the
\textsc{Matslise} package\cite{Ledoux2005} for \textsc{Matlab}. This
yields results accurate to about 14 significant digits.

\begin{figure*}
\includegraphics[width=0.45\textwidth]{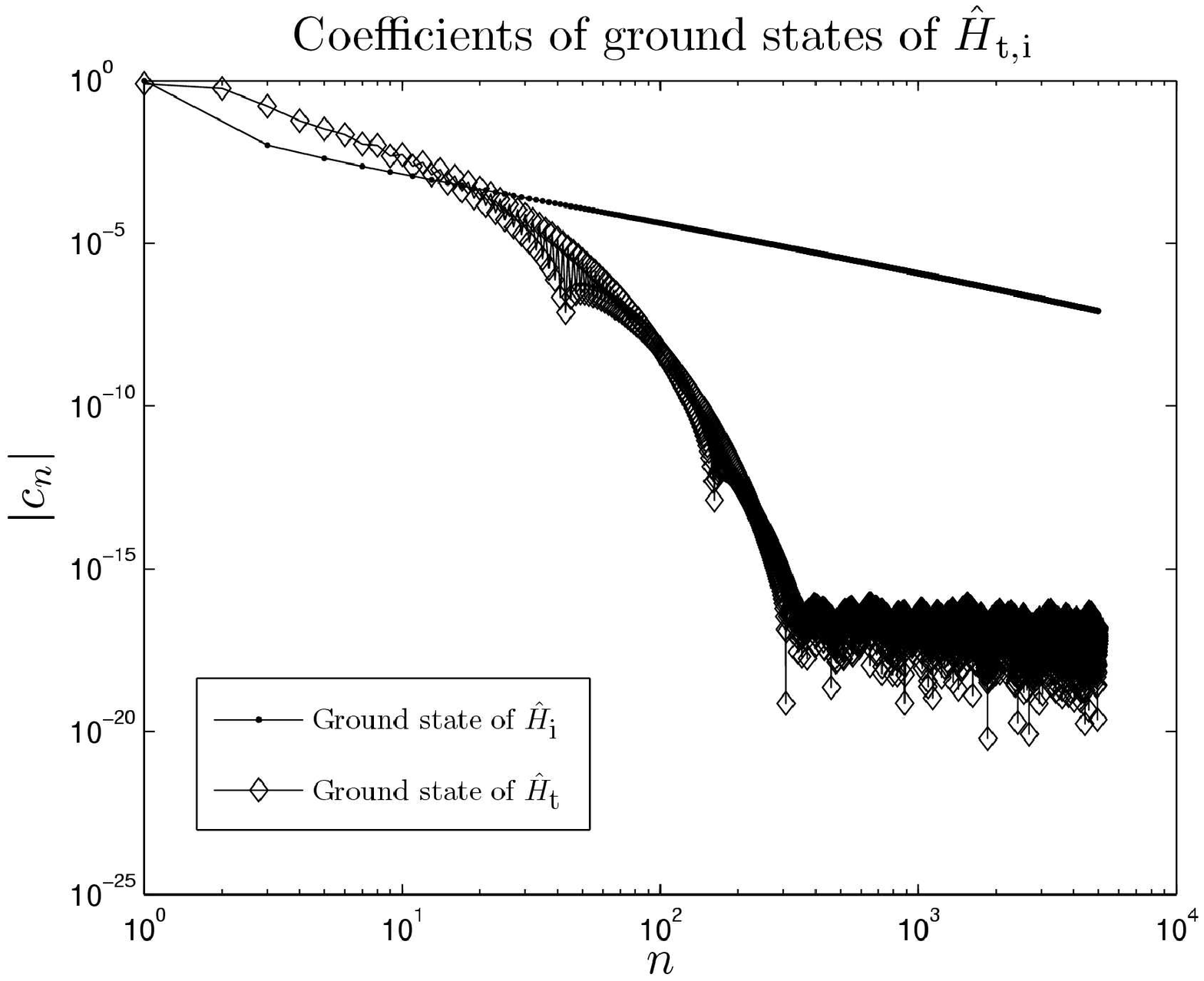}%
\includegraphics[width=0.45\textwidth]{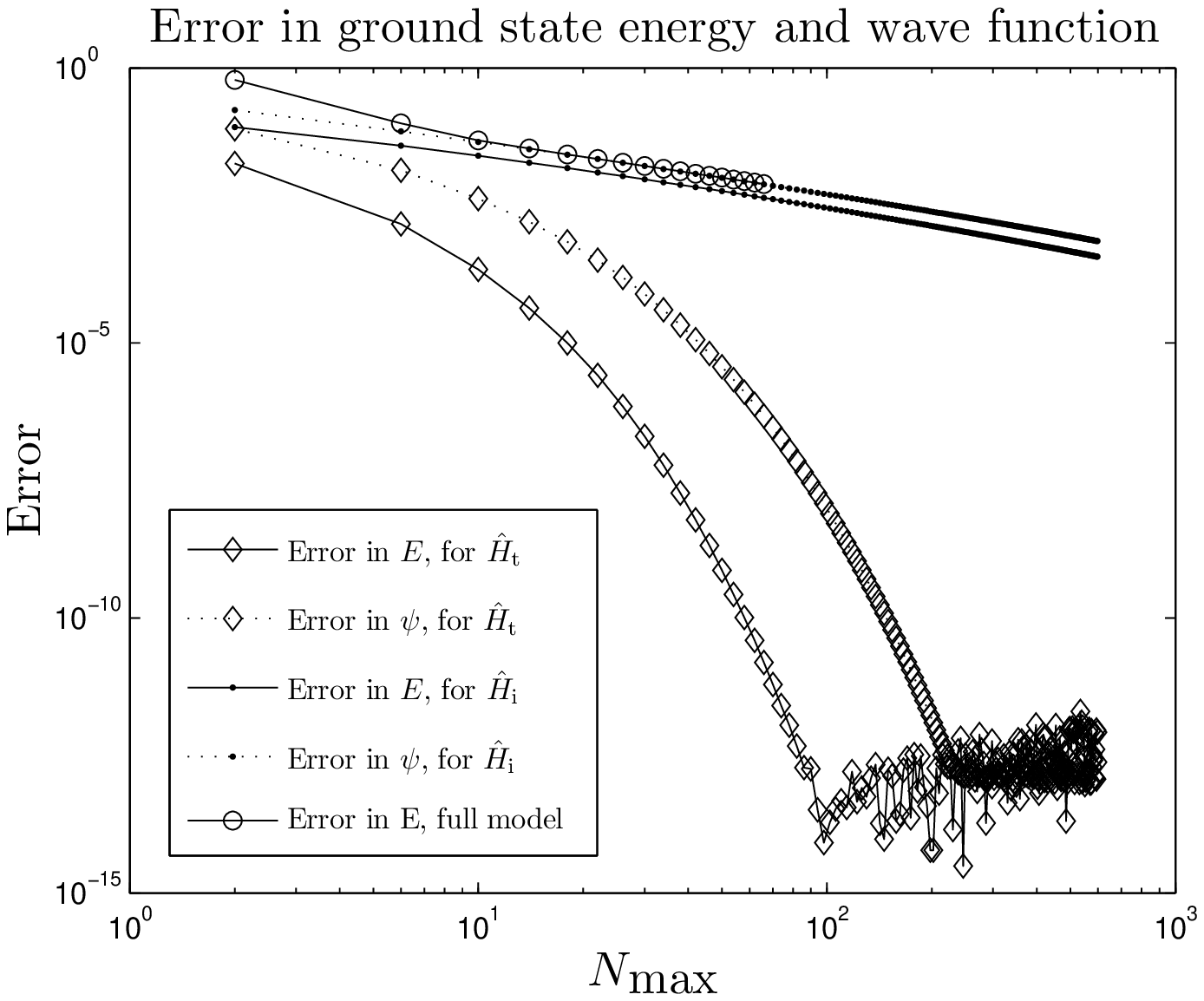}%
\caption{Left: Coefficients of the exact ground states of
  the Hamiltonians $\hat{H}_\text{t,i}$. For $\hat{H}_\text{i}$ only
  even-numbered coefficients are nonzero and thus displayed. The
  almost straight line indicates approximately $o(n^{-1.57})$ decay of
  the coefficients around $n=600$ and $o(n^{-1.73})$ around
  $n=5000$. Compare with Fig.~\ref{fig:bad-func}. For the
  $\hat{H}_\text{t}$ ground state we clearly have spectral
  convergence. Right: The error in the ground state energies and wave
  functions when using the CI method. For $\hat{H}_\text{i}$ we have $o(n^{-1.24})$ decay for
  the energy error, and $o(n^{-1.20})$ decay for the wave function
  error, both evaluated at $n=600$. For $\hat{H}_\text{t}$ we clearly
  have spectral convergence. A full two-particle CI calculation is
  superimposed, showing that the error in the interaction part of the
  Hamiltonian \eqref{eq:ham2} completely dominates. Here, the error in
  the energy is $o(N_\text{max}^{-1.02})$ at $n=70$, while for
  $\hat{H}_\text{i}$ alone, we have
  $o(N_\text{max}^{-1.01})$.\label{fig:numeric}}
\end{figure*}

In Fig.~\ref{fig:numeric} (left) the magnitude of the coefficients of
the \emph{exact} ground states alongside the ground state energy error
and wave function error (right) are graphed for each Hamiltonian,
using successively larger $N_\text{max}$. The coefficients of the
exact ground states decay according to expectations, as we clearly
have spectral convergence for the $\hat{H}_\text{t}$ ground state, and
$o(n^{-1.57})$ convergence for the $\hat{H}_\text{i}$ ground
state. 

These aspects are clearly reflected in the CI calculations. Both the
$\hat{H}_\text{t}$ ground state energy and wave function converge
spectrally with increasing $N_\text{max}$, while for
$\hat{H}_\text{i}$ we clearly have algebraic convergence. Note that
for $\hat{H}_\text{t}$, $N_\text{max} \sim
40$ yields a ground state energy accurate to $\sim 10^{-10}$, and that
such precision would require $N_\text{max} \sim 10^{12}$ for
$\hat{H}_\text{i}$, which converges only algebraically.

Intuitively, these results are easy to understand: For the trap
Hamiltonian a modest value of $N_\text{max}$ produces almost exact
results, since the exact ground state has extremely small components
outside the model space. This is not possible for the interaction
Hamiltonian, whose exact ground state is poorly approximated in the
model space alone.

If we consider the complete Hamiltonian \eqref{eq:ham2}, we now expect
the error to be dominated by the low-order convergence of the
interaction Hamiltonian eigenproblem.
Fig.~\ref{fig:numeric} also shows the error in the ground state energy
for the corresponding two-particle calculation, and the error is
indeed seen to behave identically to the $\hat{H}_\text{i}$ ground
state energy error. (That the energy error curve is almost on top of
the error in the \emph{wave function} for $\hat{H}_\text{i}$ is merely
a coincidence.)

It is clear that the non-smooth character of the potential $U$
destroys the convergence of the method. The eigenfunctions will be non-smooth,
while the basis functions are all very smooth. Of course, a non-smooth
potential $v(x)$ would destroy the convergence as well.

In this sense, we speak of a ``small perturbation $\hat{H}_1$'' if the
eigenvalues and eigenfunctions of the total Hamiltonian converge
spectrally. Otherwise, the perturbation is so strong that the very
smoothness property of the eigenfunctions vanish. In our case, even for
arbitrary small interaction strengths $\lambda$, the eigenfunctions are
non-smooth, so that the interaction is never small in the sense
defined here. On the other hand, the trap modification $v(x)$
represents a small perturbation of the harmonic oscillator if it is
smooth and rapidly decaying. This points
to the basic deficiency of the choice of h.o.~basis functions: They do
not capture the properties of the eigenfunctions.

We could overcome this problem by choosing a different set of basis
functions for the Hilbert space, and thus a different model space
$\mathcal{P}$ altogether. However, the symmetries of the h.o.~lets us
treat the interaction potential with ease by explicitly performing the
centre-of-mass transformation, a significant advantage in many-body
calculations. In our one-dimensional case, we could replace
$U(x_1-x_2)$ by a smooth potential; after all $U$ is just an
approximation somewhat randomly chosen. We would then obtain much
better results with the CI method. However, we are not willing to
trade the bare Coulomb interaction in two (or even three) dimensions
for an approximation. After all we know that the singular and
long-range nature of the interaction is essential.

We therefore propose to use \emph{effective interaction theory} known
from many-body physics to improve the accuray of CI calculations for
quantum dots. This replaces the matrix in the h.o.~basis of the
interaction term with an approximation, giving exact eigenvalues in
the case of no trap perturbation $v(x)$, regardless of the energy cut
parameter $N_\text{max}$. We cannot hope to gain spectral convergence;
the eigenfunctions are still non-smooth. However, we can increase the
algebraic convergence considerably by modifying the interaction matrix
for the given model space. This is explained in detail in the
next section.

\section{Effective Hamiltonian theory}
\label{sec:effective-interactions}

\subsection{Similarity transformation approach}

The theories of effective interactions have been, and still are, vital
ingredients in many-body physics, from quantum chemistry to nuclear
physics.\cite{helgaker,lindgren,mhj95,dickhoff,blaizot,caurier2005} In
fields like nuclear physics, due to the complicated nature of the
nuclear interactions, no exact spatial potential exists for the
interactions between nucleons. Computation of the matrix elements of
the many-body Hamiltonian then amounts to computing, for example,
successively complicated Feynman diagrams,\cite{mhj95,dickhoff}
motivating realistic yet tractable approximations such as effective
two-body interactions.  These effective interactions are in turn used
as starting points for diagonalization calculations in selected model
spaces.\cite{caurier2005,navratil1998,bruce2,navratil2000a}
Alternatively, they can be used as starting point for the resummation
of selected many-body correlations such as in coupled-cluster
theories.\cite{helgaker} In our case, it is the so-called \emph{curse
of dimensionality} that makes a direct approach unfeasible: The number
of h.o.~states needed to generate accurate energies and wave functions
grows exponentially with the number of particles in the
system. Indeed, the dimension of $\mathcal{P}$ grows as
$N_\text{max}^{Nd}/(Nd)!$

For the derivation of the effective interaction, we consider the
Hamiltonian \eqref{eq:ham2} in centre-of-mass coordinates, i.e.,
\begin{eqnarray*}
 \hat{H} &=& \hat{h}(X) + \hat{h}(x) +
v\big((X + x)/\sqrt{2}\big) +\\* & & v\big((X-x)/\sqrt{2}\big) +
U(\sqrt{2}x;\lambda,\delta). 
\end{eqnarray*}
For $v(x)\neq 0$, the Hamiltonian is clearly not separable. The
idea is then to treat $v(x_j)$ as perturbations of a system separable
in centre-of-mass coordinates; after all the trap potential is assumed
to be smooth. This new unperturbed Hamiltonian reads
\[ \hat{H}' = \hat{h}(X) + \hat{h}(x) + \hat{V}, \]
where $\hat{V}=U(\sqrt{2}x;\lambda,\delta)$, or any other
interaction in a more general setting. We wish to replace the CI
matrix of $\hat{H}'$ with a different matrix
$\hat{H}'_\text{eff}$, having the exact
eigenvalues of $\hat{H}'$, but necessarily only approximate
eigenvectors. 

The effective Hamiltonian $\hat{H}'_\text{eff}$ can be viewed as an
operator acting in the model space while embodying information about
the original interaction in the \emph{complete} space
$\mathcal{H}$. We know that this otherwise neglected part of Hilbert
space is very important if $\hat{V}$ is not small. Thus, the first
ingredient is the splitting of the Hilbert space into the model space
$\mathcal{P}=P\mathcal{H}$ and the \emph{excluded space} $\mathcal{Q}
= Q \mathcal{H} = (1-P)\mathcal{H}$. Here, $P$ is the
orthogonal projector onto the model space.

In the following, we let $N$ be the dimension of the model space
$\mathcal{P}$. There should be no danger of confusion with the number
of particles $N=2$, as this is now fixed. Moreover, we let
$\{\Phi_n\}_{n=1}^{N}$ be an orthonormal basis for $\mathcal{P}$, and
$\{\Phi_n\}_{n=N+1}^{\infty}$ be an orthonormal basis for $\mathcal{Q}$.

The second ingredient is a \emph{decoupling operator $\omega$}. It
is an operator defined by the properties
\[ P\omega = \omega Q = 0, \]
which essentially means that $\omega$ is a mapping from the model
space to the excluded space. Indeed,
\begin{eqnarray*}
  \omega &=& (P+Q)\omega(P+Q) = P\omega P+ P\omega Q + Q\omega P
  + Q\omega Q \nonumber\\* &=& Q\omega P, 
\end{eqnarray*} 
which shows that the kernel of $\omega$ includes $\mathcal{Q}$, while
the range of $\omega$ excludes $\mathcal{P}$, i.e., that $\omega$ acts
only on states in $\mathcal{P}$ and yields only states in
$\mathcal{Q}$.

The effective Hamiltonian $\hat{H}_\text{eff} = P[\hat{h}(x)+\hat{h}(X)]P +
\hat{V}_\text{eff}$, where $\hat{V}_\text{eff}$ is the effective interaction, is
given by the similarity transformation\cite{Suzuki1994}
\begin{equation}
    \hat{H}_\text{eff} = Pe^{-z}\hat{H}e^{z}P,
    \label{eq:sim} 
\end{equation} 
where $z = \operatorname{artanh}(\omega-\omega^\dag)$.
The key point is that $e^{z}$ is a unitary operator with $(e^{z})^{-1} =
e^{-z}$, so that the $N$ eigenvalues of $\hat{H}'_\text{eff}$ are actually
eigenvalues of $\hat{H}'$.

In order to generate a well-defined effective Hamiltonian, we must
define $\omega=Q\omega P$ properly. The approach of Suzuki and
collaborators\cite{Suzuki1982,suzuki2,Suzuki1995,Suzuki1994} is
simple: Select an orthonormal set of vectors
$\{\chi_n\}_{n=1}^N$. These can be some eigenvectors of $\hat{H}'$ we
wish to include. Assume that $\{P\chi_n\}_{n=1}^N$ is a basis for the
model space, i.e., that for any $n\leq N$ we can write
\[ \Phi_n = \sum_{m=1}^N a_{n,m} P\chi_m \]
for some constants $a_{n,m}$. We then define $\omega$ by
\[ \omega P\chi_n := Q\chi_n, \quad n=1,\ldots,N. \]
Observe that $\omega$ defined in this way is an operator that
reconstructs the excluded space components of $\chi_n$ given its model space
components, thereby indeed embodying information about the Hamiltonian
acting on the excluded space.

Using the decoupling properties of $\omega$ we quickly calculate
\[ \omega \Phi_n = Q\omega P\Phi_n = Q \omega \sum_{m=1}^N a_{n,m}
\chi_m, \quad n =1,\ldots, N \]
and hence for any $n'>N$ we have
\[ \langle \Phi_{n'} , \omega \Phi_n \rangle = \sum_{m=1}^N a_{n,m}
\langle \Phi_{n'}, \chi_m \rangle, \] yielding all the non-zero matrix
elements of $\omega$.

As for the vectors $\chi_n$, we do not know \emph{a priori} the exact
eigenfunctions of $\hat{H}'$, of course. Hence, we cannot find
$\hat{H}'_\text{eff}$ exactly. The usual way to find the eigenvalues is
to solve a much larger problem with $N'>N$ and then assume that these
eigenvalues are ``exact''. The reason why this is possible at all is
that our Hamiltonian $\hat{H}'$ is separable, and therefore easier to
solve.  However, we have seen that this is a bad method: Indeed, one
needs a matrix dimension of about $10^{10}$ to obtain about 10
significant digits. Therefore we instead reuse the aforementioned
constant reference potential method to obtain eigenfunctions and
eigenvectors accurate to machine precision.

Which eigenvectors of $\hat{H}'$ do we wish to include? Intuitively,
the first choice would be the lowest $N$ eigenvectors. However, simply
ordering the eigenvalues ``by value'' is not what we want
here. Observe that $\hat{H}'$ is block diagonal, and that the model
space contains $N_\text{max}+1$ blocks of sizes 1 through $N_\text{max}+1$. If we
look at the \emph{exact} eigenvalues, we know that they have the structure
\[ E_{n,m} = (n + 1/2) + \epsilon_m, \]
where $n$ is the block number and $\epsilon_m$ are the eigenvalues of
$\hat{H}_\text{i}$, see Eq.~\eqref{eq:inter-ham}. But it is easy to
see that the large-scale diagonalization eigenvalues do \emph{not}
have this structure -- we only obtain this in the limit
$N_\text{max}\rightarrow\infty$. Therefore we choose the eigenvectors
corresponding to the $N$ eigenvalues $E_{n,m}$, $n+m\leq
N_\text{max}$, thereby achieving this structure in the eigenvalues of
$\hat{H}'_\text{eff}$.

In general, we wish to incorporate the symmetries of $\hat{H}'$ into
the effective Hamiltonian $\hat{H}'_\text{eff}$. In this case, it was
the separability and even eigenvalue spacing we wished to
reproduce. In Sec.~\ref{sec:discussion} we treat the
two-dimensional Coulomb problem similarly.

\subsection{Numerical experiments with effective interactions}

The eigenvectors of the Hamiltonian $\hat{H}'$ differ from those of
the the effective Hamiltonian $\hat{H}'_\text{eff}$. In this section,
we first make a qualitative comparison between the ground states of
each Hamiltonian. We then turn to a numerical study of the error in
the CI method when using the effective interaction in a the model
problem.

\begin{figure*}
\includegraphics[width=0.45\textwidth]{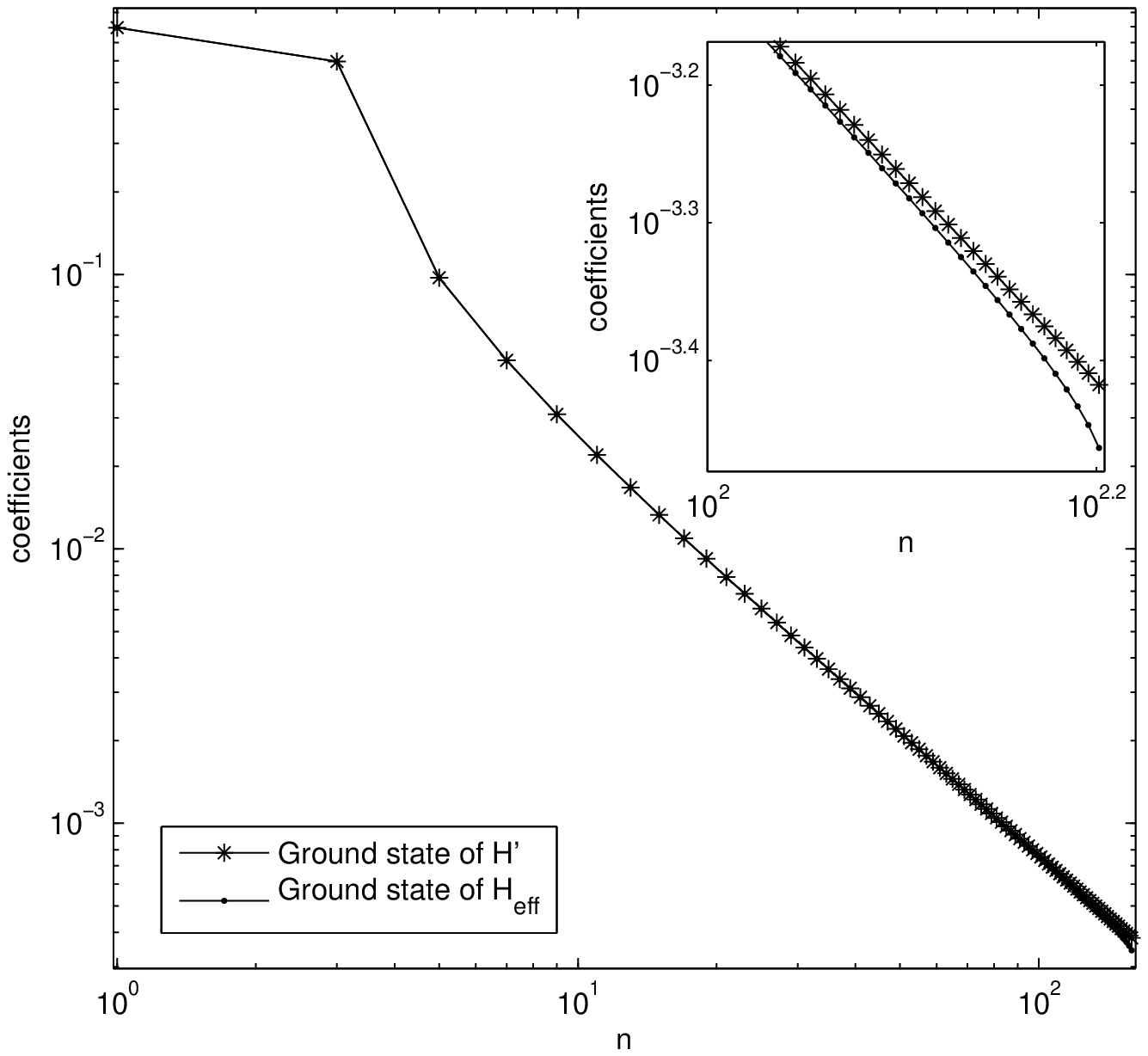}%
\includegraphics[width=0.45\textwidth]{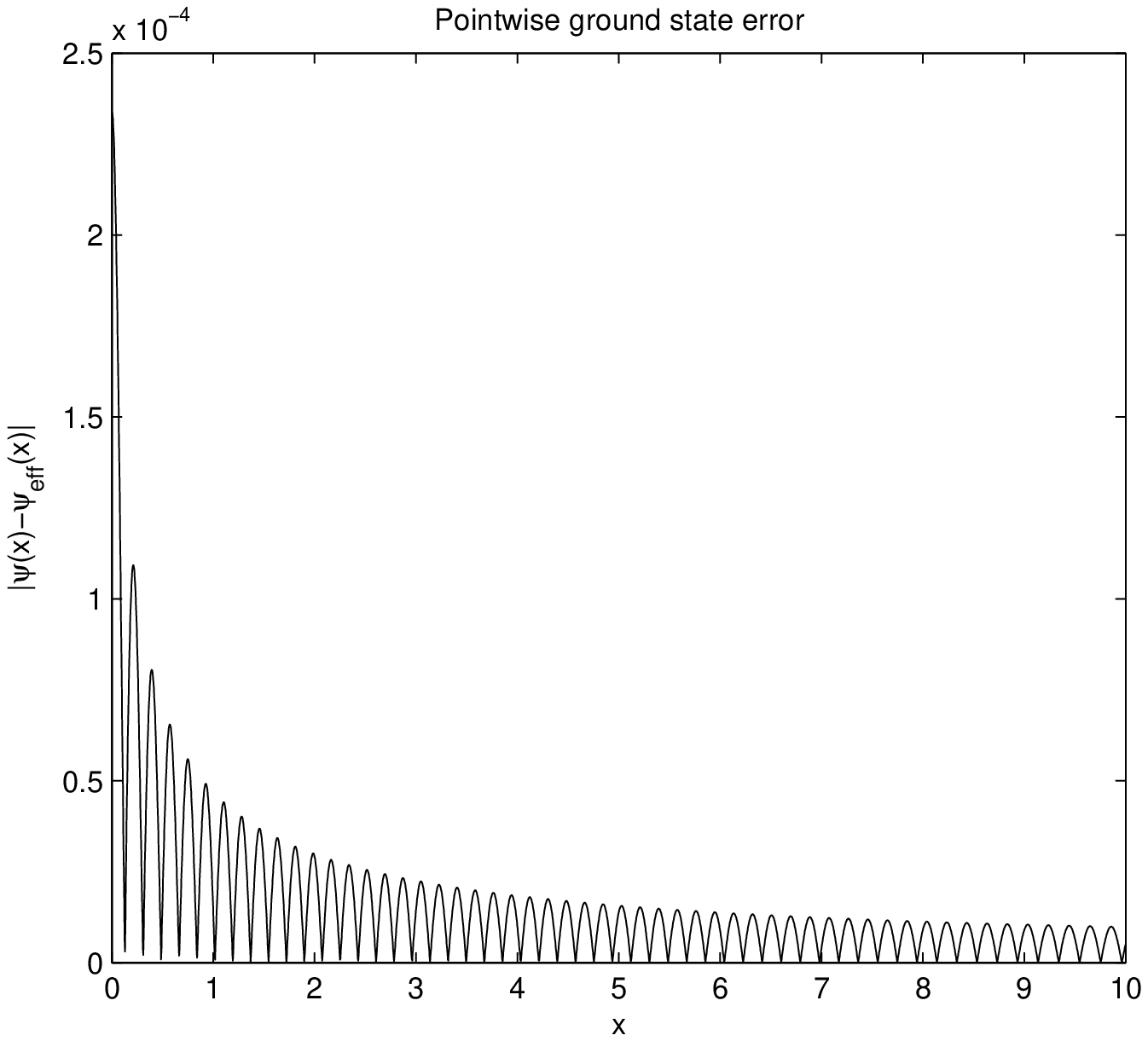}
\caption{Left: Plot of ground state coefficients of $\hat{H}'$
  and $\hat{H}_\text{eff}$. Right: Pointwise error (in relative
  coordinate $x$) of effective Hamiltonian ground
  state $\psi_\text{eff}(x)$}\label{fig:effective-eigenvectors}
\end{figure*}

Recall that the ground state eigenvectors are on the form
\[ \Psi(X, x) = \phi_0(X)\psi(x) = \phi_0(X)\sum_{n=0}^{\infty} c_n
\phi_n(x). \] For $\hat{H}'_\text{eff}$, $c_n=0$ for all
$n>N_\text{max}$, so that the excluded space-part of the error
concides with the excluded space-part of the exact ground state.  In
Fig.~\ref{fig:effective-eigenvectors} the coefficients $c_n$ for both
$\hat{H}'$ and $\hat{H}'_\text{eff}$ are displayed. The pointwise
error is also plotted, and the largest values are seen to be around
$x=0$. This is expected since $U(\sqrt{2}x;\lambda,\delta)$ and the
exact ground state is non-smooth there. Notice the slow spatial decay
of the error, intuitively explained by the slow decay of the Coulomb
interaction.

We now turn to a simulation of the full two-particle Hamiltonian
\eqref{eq:ham2}, and compare the decay of the ground state energy
error with and without the effective interaction. Thus, we perform two
simulations with Hamiltonians
\begin{eqnarray*}
  \hat{H} &=& \hat{H}' + v(x_1) + v(x_2) \nonumber\\*
  &=& \hat{h}(x_1) + \hat{h}(x_2) + v(x_1) + v(x_2) +
\hat{T}\hat{V}\hat{T}^\dag
\end{eqnarray*}
and
\begin{eqnarray*}
  \hat{H}_\text{eff} &=& \hat{H}'_\text{eff} + v(x_1) + v(x_2) \nonumber\\*
  &=& \hat{h}(x_1) + \hat{h}(x_2) + v(x_1) + v(x_2) +
\hat{T}\hat{V}_\text{eff}\hat{T}^\dag,
\end{eqnarray*}
respectively, where $\hat{T}$ is the centre-of-mass transformation,
cf.~Eq.~\eqref{eq:trafo}.

We remark that the new Hamiltonian matrix has the \emph{same
structure} as the original matrix. It is only the values of the
interaction matrix elements that are changed. Hence, the new scheme
has the same complexity as the CI method if we disregard the
computation of $\hat{V}_\text{eff}$, which is a one-time calculation
of low complexity.

\begin{figure}
\includegraphics[width=0.5\textwidth]{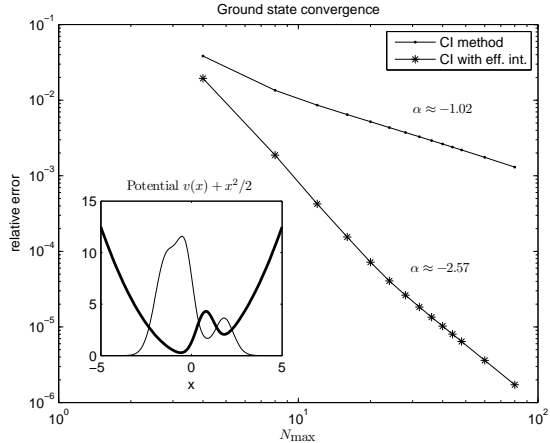}%
\caption{Ground state energy relative error for a two-particle simulation using the
  confinement potential $V(x) = x^2/2 + 4\exp(-2(x-0.75)^2)$. For the
  CI method without effective interactions,
  we obtain $\alpha\approx -1.02$, while the effective interactions
  gives $\alpha\approx -2.57$. The electron density is superimposed on
  the potential plot.} 
\label{fig:full-dot-sim}
\end{figure}

The results are striking: In Fig.~\ref{fig:full-dot-sim} we see that
the ground state error decays as $O(N_\text{max}^{-2.57})$, compared
to 
$O(N_\text{max}^{-0.95})$ for the original CI method. For
$N_\text{max} = 40$, the CI relative error is $\Delta E/E_0
\approx 2.6\cdot 10^{-3}$, while for the effective interaction
approach $\Delta E/E_0 \approx 1.0\cdot 10^{-5}$, a considerable gain.

The ground state energy $E_0$ used for computing the errors were computed
using extrapolation of the results. 

We comment that $N_\text{max}\sim 40$ is the practical limit on a
single desktop computer for a two-dimensional two-particle
simulation. Adding more particles further restricts this limit,
emphasizing the importance of the gain achieved in the relative
error.

In a more systematical treatment, we computed the error decay
coefficient $\alpha$ for a range of trap potentials $x^2 +
A\exp(-2(x-\mu)^2)$, where we vary $A$ and $\mu$ to create single and
double-well potentials. In most cases we could estimate $\alpha$
successfully. For low values of $\mu$, i.e., near-symmetric wells, the
parameter estimation was difficult in the effective interaction case
due to very quick convergence of the energy. The CI calculations also
converged quicker in this case. Intuitively this is so because the two
electrons are far apart in this configuration.

\begin{figure*}
\includegraphics[width=0.45\textwidth]{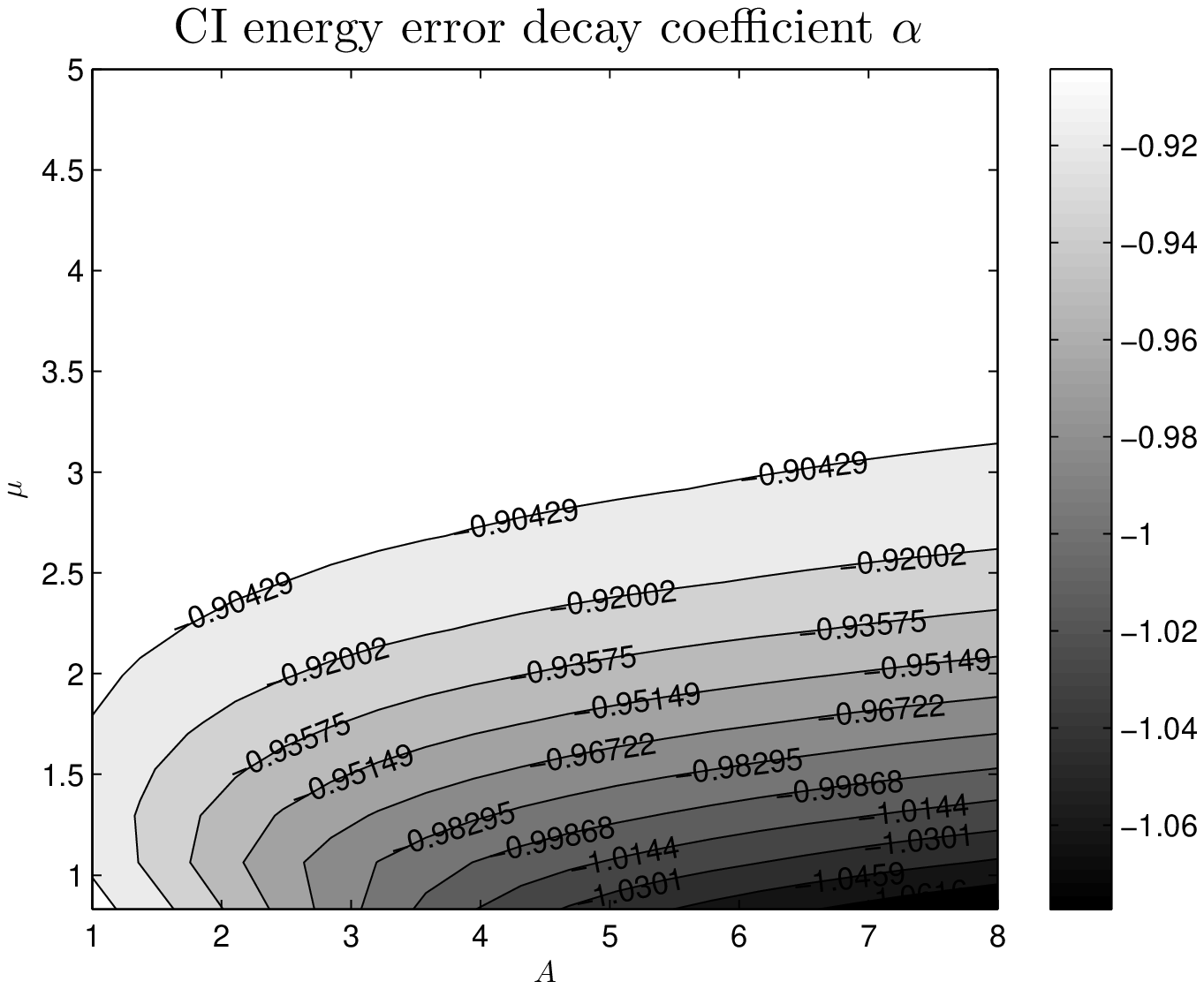}%
\includegraphics[width=0.45\textwidth]{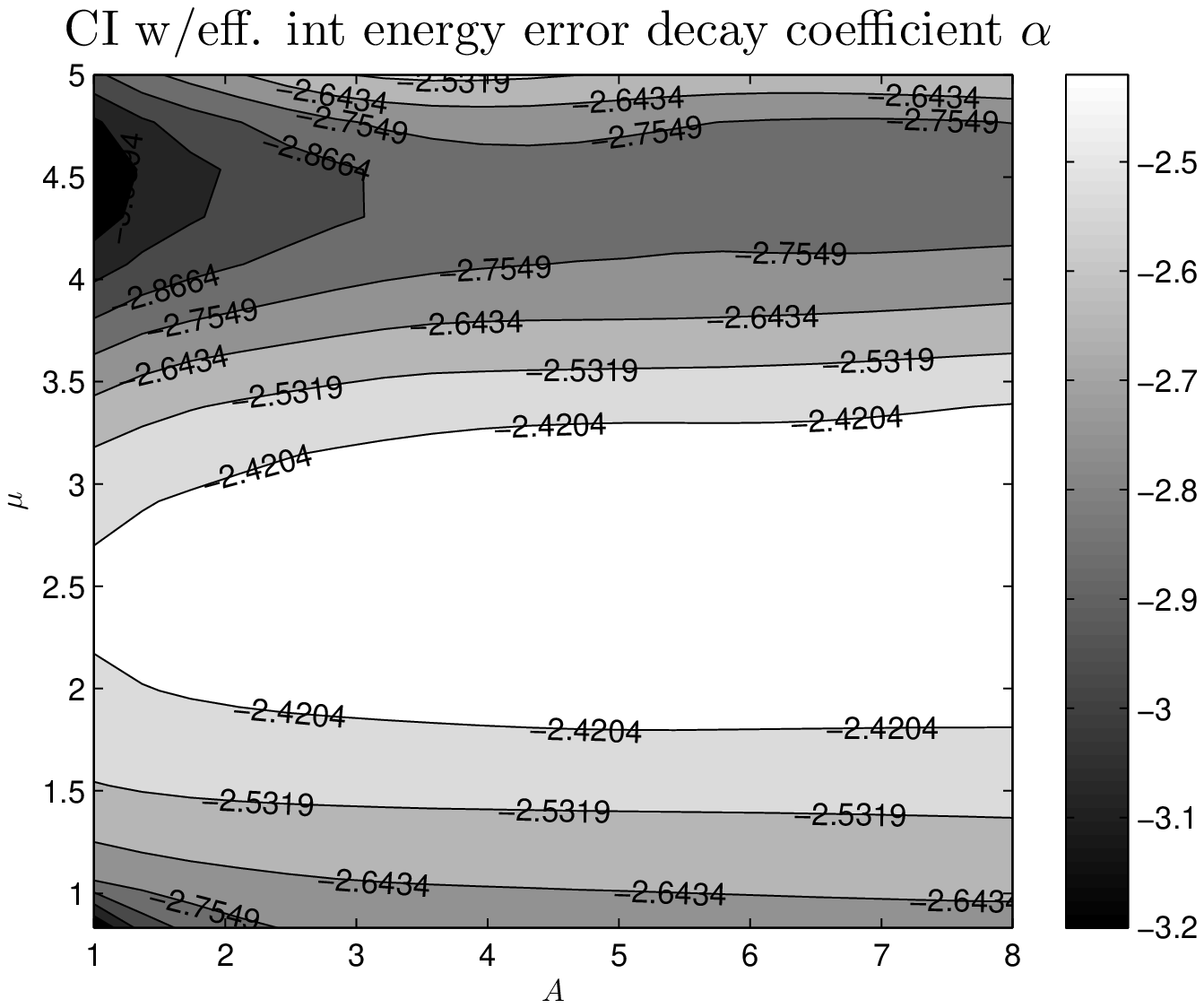}%
\caption{Estimates of $\alpha$ for CI calcilations with (right) and
  without (left) effective interactions.}\label{fig:alpha}
\end{figure*}

The results indicate that at $N_\text{max} = 60$ we have 
\[ \alpha = -0.96 \pm 0.04 \quad\text{for $\hat{H}$} \]
and 
\[ \alpha = -2.6 \pm 0.2 \quad\text{for $\hat{H}_\text{eff}$} \]
for the chosen model. Here, $0.6\leq\mu\leq 1.8$ and $2.9 \leq A \leq
4.7$ and all the fits were successful. In Fig.~\ref{fig:alpha}
contour plots of the obtained results are shown. For the shown range,
results were unambiguous.

These numerical results clearly indicate that the effective interaction
approach will gain valuable numerical precision over the original CI
method in general; in fact we have gained nearly two orders of
magnitude in the decay rate of the error.

\section{Discussion and outlook}
\label{sec:discussion}

\subsection{Generalizations}

One-dimensional quantum dot models are of limited value in
themselves. However, as claimed in the Introduction, the analysis and
experiments performed in this article are valid also in
higher-dimensional systems.

Consider two particles in two dimensions. Let $\hat{h}(\vec{r})$ be
the two-dimensional harmonic oscillator Hamiltonian (we omit the
superscript in Eq.~\eqref{eq:ho-ham} for brevity), and let the
quantum dot Hamiltonian be given by
\[
  \hat{H} = \hat{H}' + v(\vec{r}_1) + v(\vec{r}_2),
\]  
where 
\[ \hat{H}' = \hat{h}(\vec{r}_1) + \hat{h}(\vec{r}_2) +
\frac{\lambda}{\|\vec{r}_1-\vec{r}_2\|}.
\]
The normalized centre-of-mass and relative coordinates are defined by
\[ \vec{R} = \frac{\vec{r}_1 + \vec{r}_2}{\sqrt{2}} \quad \text{and}
\quad \vec{r} = \frac{\vec{r}_1 - \vec{r}_2}{\sqrt{2}}, \]
respectively, which gives
\[ \hat{H}' = \hat{h}(\vec{R}) + \hat{h}(\vec{r}) + \frac{\lambda}{\sqrt{2}\|\vec{r}\|}. \]
The h.o.~eigenfunctions in polar coordinates are given
by\cite{Rontani2006}
\[ \Phi_{n,m}(r,\theta) \propto e^{im\theta} r^{|m|}
L_n^{|m|}(r^2)e^{-r^2/2}, \] and the corresponding eigenvalues are
$2n+|m|+1$. Now, $\hat{H}'$ is further separable in polar coordinates,
yielding a single radial eigenvalue equation to solve, analoguous to
the single one-dimensional eigenvalue equation of $\hat{H}_\text{i}$
in Eq.~\eqref{eq:inter-ham}.

The eigenvalues of $\hat{H}'$ have the structure
\[ E_{n',m',n,m} = 2n'+|m'| + 1 + \epsilon_{n,m}, \]
where $(n',m')$ and $(n,m)$ are the centre-of-mass and relative
coordinate quantum numbers, respectively. Again, the degeneracy
structure and even spacing of the eigenvalues are destroyed in the CI
approach, and we wish to regain it with the effective interaction. We
then choose the eigenvectors corresponding to the quantum numbers
\[ 2n' + |m'| + 2n + m \leq N_\text{max} \]
to build our effective Hamiltonian $\hat{H}'_\text{eff}$.

Let us also mention, that the exact eigenvectors $\Psi_{n',m',n,m}$
are non-smooth due to the $1/r$-singularity of the Coulomb
interaction. The approximation properties of the Hermite functions are
then directly applicable as before, when we expand the eigenfunctions
in h.o.~basis functions. Hence, the configuration-interaction method
will converge slowly also in the two-dimensional case. It is good
reason to believe that effective interaction experiments will
yield similarly positive results with respect to convergence
improvement.

Clearly, the above procedure is applicable to three-dimensional
problems as well. The operator $\hat{H}'$ is separable and we obtain a
single non-trivial radial equation, and thus we may apply our
effective Hamiltonian procedure. The exact eigenvalues will have the
structure
\[ E_{n',l',m',n,l,m} = 2n' + l' + \frac{3}{2} + \epsilon_{n,l,m}, \]
on which we base the choice of the effective Hamiltonian eigenvectors
as before.

The effective interaction approach to the configuration-interaction
calculations is easily extended to a many-particle problem, whose
Hamiltonian is given by Eq.~\eqref{eq:big-hamiltonian}. The form of
the Hamiltonian contains only interactions between pairs of particles,
and $\hat{V}_\text{eff}$ as defined in
Sec.~\ref{sec:effective-interactions} can simply replace these terms.

\subsection{Outlook}

A theoretical understanding of the behavior of many-body systems is a
great challenge and provides fundamental insights into quantum
mechanical studies, as well as offering potential areas of
applications. However, apart from some few analytically solvable
problems, the typical absence of an exactly solvable contribution to
the many-particle Hamiltonian means that we need reliable numerical
many-body methods. These methods should allow for controlled
expansions and provide a calculational scheme which accounts for
successive many-body corrections in a systematic way.  Typical
examples of popular many-body methods are coupled-cluster
methods,\cite{bartlett81,helgaker,Wloch2005} various types of Monte
Carlo methods,\cite{Pudliner1997,ceperley1995,mc3} perturbative
expansions,\cite{lindgren,mhj95} Green's function
methods,\cite{dickhoff,blaizot} the density-matrix renormalization
group\cite{white1992,schollwock2005} and large-scale diagonalization
methods such as the CI method considered here.

In a forthcoming article, we will apply the similarity transformed effective
interaction theory to a two-dimensional system, and also extend the
results to many-body situations. Application of other methods, such as
coupled-cluster calculations, are also an interesting approach, and
can give further refinements on the convergence, as well as gaining
insight into the behaviour of the numerical methods in general.

The study of this effective Hamiltonian is interesting from a
many-body point of view: The effective two-body force is built from a
two-particle system. The effective two-body interaction derived from
an $N$-body system, however, is not necessarly the same. Intuitively,
one can think of the former approach as neglecting interactions and
scattering between three or more two particles at a time.  In nuclear
physics, such three-body correlations are non-negligible and improve
the convergence in terms of the number of harmonic oscillator
shells.\cite{Navratil2003} Our hope is that such
interactions are much less important for Coulomb systems.

Moreover, as mentioned in the Introduction, accurate determination of
eigenvalues is essential for simulations of quantum dots in the time
domain. Armed with the accuracy provided by the effective
interactions, we may commence interesting studies of quantum dots
interacting with their environment.

\subsection{Conclusion}

We have mathematically and numerically investigated the properties of
the configuration-interaction method, or ``exact diagonalization
method'', by using results from the theory of Hermite series.  The
importance of the properties of the trap and interaction potentials is
stressed: Non-smooth potentials severely hampers the numerical
properties of the method, while smooth potentials yields exact results
with reasonable computing resources. On the other hand, the h.o.~basis
is very well suited due to the symmetries under orthogonal coordinate
changes. 

In our numerical experiments, we have demonstrated that for a simple
one-dimensional quantum dot with a smooth trap, the use of similarity
transformed effective interactions can significantly reduce the error
in the configuration-interaction calculations due to the non-smooth
interaction, while not increasing the complexity of the
algorithm. This error reduction can be crucial for many-body
simulations, for which the number of harmonic oscillator shells is
very modest.

\end{document}